\documentclass[aps,prb,reprint]{revtex4-1}
\usepackage{amsmath}
\usepackage{txfonts}
\usepackage{amssymb}
\usepackage{graphicx}% Include figure files
\usepackage{bm}% bold math
\usepackage{color}
\usepackage{pdfpages}

\newcommand{\Rmnum}[1]{\expandafter\@slowromancap\romannumeral #1@}

\newcommand{\be}{\begin{equation}}
\newcommand{\ee}{\end{equation}}
\newcommand{\bfig}{\begin{figure}}
\newcommand{\efig}{\end{figure}}

\begin{document}

\title{How much better are InGaN/GaN nanodisks than quantum wells -- oscillator strength enhancement and changes in optical properties}

\author{Lei Zhang}
\affiliation{Department of Physics, University of Michigan, 450 Church Street, Ann Arbor, MI 48109, USA}
\author{Leung-Kway Lee}
\affiliation{Department of Electrical Engineering and Computer Science, University of Michigan, 1301 Beal Ave., Ann Arbor, MI 48109, USA}
\author{Chu-Hsiang Teng}
\affiliation{Department of Electrical Engineering and Computer Science, University of Michigan, 1301 Beal Ave., Ann Arbor, MI 48109, USA}
\author{Tyler A. Hill}
\affiliation{Department of Physics, University of Michigan, 450 Church Street, Ann Arbor, MI 48109, USA}
\author{Pei-Cheng Ku}
\affiliation{Department of Electrical Engineering and Computer Science, University of Michigan, 1301 Beal Ave., Ann Arbor, MI 48109, USA}
\author{Hui Deng}
\email{Correspondence should be sent to dengh@umich.edu regarding optical measurements and to peicheng@umich.edu regarding materials synthesis.}
\affiliation{Department of Physics, University of Michigan, 450 Church Street, Ann Arbor, MI 48109, USA}

\begin{abstract}
\label{Abstract}
We show over 100-fold enhancement of the exciton oscillator strength as the diameter of an InGaN nanodisk in a GaN nanopillar is reduced from a few micrometers to less than 40 nm, corresponding to the quantum dot limit. The enhancement results from significant strain relaxation in nanodisks less than 100 nm in diameter. Meanwhile, the radiative decay rate is only improved by 10 folds due to strong reduction of the local density of photon states in small nanodisks. Further increase in the radiative decay rate can be achieved by engineering the local density of photon states, such as adding a dielectric coating.
\end{abstract}

\date{\today}

\maketitle

\label{Introduction}
InGaN/GaN quantum wells (QWs), with a bandgap tunable over the full visible spectral range, play a vital role in high efficiency visible light emitting diodes (LEDs) and laser diodes\cite{Nakamura1998}. InGaN/GaN quantum dots (QDs) also hold the promise as high temperature quantum photonic devices \cite{Kako2006b, Deshpande2013}. Unfortunately, a large, strain-induced electric field in III-N heterostructures often severely suppresses the oscillator strength of the exciton and, hence, the radiative decay rate and the internal quantum efficiency (IQE). Since strain is relaxed near free surfaces, nanodisks (NDs) in nanowires, which has a large surface-to-volume ratio, have been widely considered as a promising solution for improving the IQE of InGaN/GaN photonic devices \cite{Kawakami2006, Holmes2009, Holmes2011a, Hsueh2005, Chen2006a, Kawakami2010, Ramesh2010, Lee2011}. The accompanying improvement in the radiative decay rate is also important for realizing ultrafast single-photon sources using InGaN/GaN QDs\cite{Zhang2013b, Deshpande2013}.

Despite the observation of enhancements in photoluminescence (PL) intensity and decay rate in InGaN/GaN NDs\cite{He2005, Kawakami2006, Ramesh2010}, the size-dependent behavior of IQE $\eta_{int}$ and radiative decay rate $\gamma_r$, mostly affected by the emitter's oscillator strength $f_{os}$, remains unclear. Experimental studies using top-down nanopillars reported up to ten-fold enhancement in emission intensity only in nanopillars larger than 100~nm in diameter\cite{Ramesh2010}. However, theoretical studies predicted that strain relaxation is most significant only in the region $<20$~nm from the sidewall, suggesting significant improvement of IQE could be possible in small disks $\sim 40$~nm in size\cite{Kawakami2010}. Furthermore, no comparisons so far have taken into account the influence of external optical efficiencies on the measured PL intensity, such as input laser absorption efficiency $\eta_{abs}$, local density of photon states (LDPS) $\rho$\cite{Purcell1946, Bleuse2011} and emission collection efficiency $\eta_{col}$\cite{Claudon2010}, which hinders the extraction of $\eta_{int}$, $\gamma_r$ as well as $f_{os}$. 

In this work, we systematically compare the PL properties of NDs with diameters ranging from 15~nm to 2~$\mu$m. We first present two NDs at the two ends of the diameter range, highlighting their fundamentally different spectral responses to the excitation-induced carrier-screening of piezoelectric fields. We then examine how PL energy, decay time and intensity changes continuously with ND diameter reduction, from which we extract the enhancement of $f_{os}$, $\gamma_r$, and $\eta_{int}$ through careful analysis of various external optical efficiencies.

\label{Sample and Optics}
The sample we used was grown by metalorganic chemical vapor deposition (MOCVD) on a sapphire (0001) substrate. A single In$_{0.15}$Ga$_{0.85}$N QW of 3~nm thickness was sandwiched between a 10~nm thick GaN layer on the top and a 1.5~$\mu$m thick GaN layer at the bottom. There is no intentional doping in any of the layers, but there may exist an unintentional n-doping of $\sim1\times10^{17}$~cm$^{-3}$. The planar 2D structure was patterned via electron beam lithography (EBL) and subsequently etched into nanopillars of 120~nm in height by inductively-coupled plasma reactive-ion etching (ICP-RIE). The finished device contains multiple $6\times6$ arrays of nanopillars, each nanopillar containing a single InGaN ND. The separation between nanopillars in each array is 5~$\mu$m, large enough for studying single-nanopillar properties by the isolation of single nanopillars with our confocal microscopy setup. A 390~nm laser obtained from frequency doubling a 780~nm Ti:Sapphire laser with a 150~fs pulse width and an 80~MHz repetition rate was focused from an angle 55~$^\circ$ apart from normal direction onto a spot of 50~$\mu$m in diameter to excite the sample. The PL was collected using an objective lens with a 40$\times$ magnification and a 0.6 numerical aperture (NA) from the normal direction. All measurements were done at 10~K temperature. Detailed sample fabrication process and the optical setup were similar to those described in our previous work \cite{Lee2011, Zhang2013b}.

\label{QW vs. QD}
\begin{figure*}
	\includegraphics{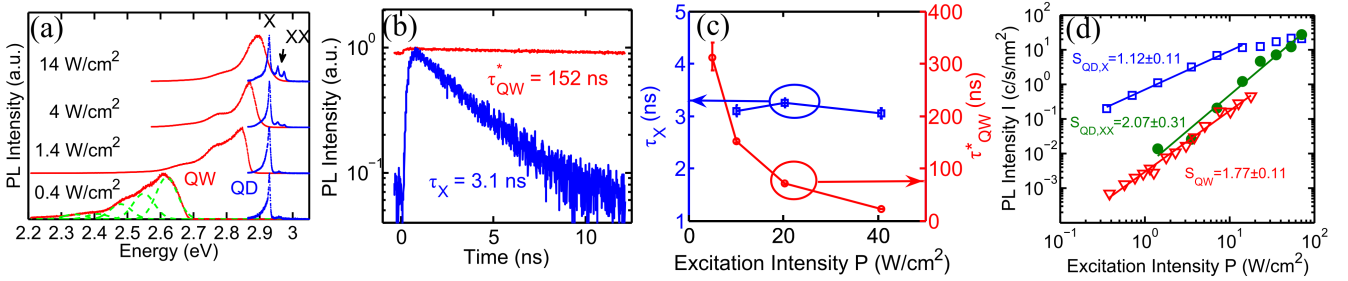}
  \caption{
(a) The PL spectra from a QW-ND with a diameter of 2~$\mu$m (red lines) and a QD-ND with a diameter of 20~nm (blue lines) at excitation intensities $P = 0.4$, 1.4, 4 and 14~W/cm$^2$. The spectra are normalized to their maxima and vertically shifted to ease the comparison.
The spectra of the QW-ND is broadened by optical-phonon replicas as illustrated in dashed green lines.
%The major emission peak X of the QD PL is at $E=2.93$~eV insensitive to $P$. At higher excitation intensities, another peak XX emerges at a higher energy $E= 2.95$~eV.
(b) The TRPL decay traces of QW-ND (red) and QD-ND's X peak (blue) measured at $P = 10$~W/cm$^2$.
(c) The mono-exponential decay time of the X peak of the QD-ND $\tau_X$ (blue square) and the decay time obtained from mono-exponential fit of the initial 12.5~ns decay trace of the QW-ND $\tau^*_{QW}$ (red triangle) at various excitation intensities.
(d) The integrated PL intensity of the X (blue square) and XX (green circle) peaks from the QD-ND and the overall integrated PL intensity, including the optical-phonon replicas, from the QW-ND (red triangle). All the intensity values $I$ are normalized by the lateral area of the InGaN layers and hence have a unit of count per second per nanometer square (c/s/nm$^2$). A linear fit of $log(I)$ vs. $log(P)$ shows that the intensities of the X and XX peaks of the QD-ND, and the spectrally integrated intensity of the QW-ND are proportional to $P^{1.12}$, $P^{2.07}$ and $P^{1.77}$, respectively.}
  \label{fig:pl_power_depend}
\end{figure*}

We first compared the optical properties of a large and a small NDs via excitation-intensity-dependent PL study. The large ND is $2~\mu$m in diameter and fully strained; it corresponds to the QW-regime and we call it QW-ND. The small ND is 15~nm in diameter and should have the least amount of strain. It was found to be QD-like and will be called QD-ND in this paper.

The emission from the QW-ND is strongly influenced by the strain in the InGaN layer, common to III-N QWs grown on $c$-plane\cite{Kawakami2006, Jarjour2007a}. At low excitation intensities, as shown in Fig.~\ref{fig:pl_power_depend}(a), the PL spectra are hundreds of meV broad with pronounced optical-phonon replicas due to the strain-enhanced wavefunction overlapping between electrons, holes and the longitudinal interface optical-phonon modes\cite{Chen2007b, Renwick2012}. The time-resolved PL (TRPL) decay of the QW-ND is very slow (Fig.~\ref{fig:pl_power_depend}(b)). In order to quantify the decay time of the QW-ND, we used a mono-exponential function to fit the initial 12.5~ns decay curve and got a $1/e$ initial decay time of $\tau^*_{QW} = 152$~ns. Such slow TRPL decay indicates a weak oscillator strength $f_{os}$ due to large strain-induced electric fields. Note, however, that it is expected that a strained InGaN QW has a stretched exponential decay trace due to the reduction of the carrier density and, thus, the increase of the piezo-electric field over time, which is not obvious in Fig.~\ref{fig:pl_power_depend}(b) due to the short time window.

With increasing excitation intensity $P$, optical-phonon replicas are reduced, the PL peak energy $E$ blueshifts by more than 300~meV (Fig.~\ref{fig:pl_power_depend}(a)), and the decay time of the emission shortens (Fig.~\ref{fig:pl_power_depend}(c)) by orders of magnitude. These can all be qualitatively explained by the screening of the built-in field with increased photo-carrier density.

The integrated PL intensity $I$, including optical-phonon replicas, increases super-linearly with the excitation intensity $P$ as $I\propto P^{1.77}$ (Fig.~\ref{fig:pl_power_depend}(d)). This is because non-radiative recombination processes either have little dependence on carrier density, such as for the Shockley-Hall-Read process\cite{Shockley1952}, or have negligible contributions over the range of excitation intensities investigated here, such as for the Auger recombination\cite{Iveland2013}. Therefore, at higher excitation intensities, an enhanced radiative decay results in an improved IQE, manifested as the super-linear increase of the PL intensity. Assuming that the photo-carrier generation rate is proportional to the excitation intensity $P$, it immediately follows that the IQE $\eta_{int}$ increases with $P$ as $\eta_{int}\propto I/P \propto P^{0.77}$.

In contrast to the QW-ND, the emission from the QD-ND show distinct properties. As shown in Fig.~\ref{fig:pl_power_depend}(a), at low excitation intensities of $P<10~$W/cm$^2$ the PL of the QD-ND is dominated by a single peak labeled as X at 2.925~eV. The mono-exponential decay time of the X emission $\tau_X = 3.1$~ns is much shorter than $\tau^*_{QW}$,  suggesting that the built-in electric field is weaker in the QD-ND than in the QW-ND. At high excitation intensities of $P>10~$W/cm$^2$, two additional peaks at higher-energies 2.955~eV (XX) and 2.970~eV (possibly charged XX) appear (Fig.~\ref{fig:pl_power_depend}(a)). The integrated intensity of the X-peak increases linearly with $P$ up to the saturation intensity of $\sim 15$~W/cm$^2$, while the intensity of the XX-peak increases quadratically with $P$ (Fig.~\ref{fig:pl_power_depend}(d)). Hence we assign peak X to the single exciton emission, and peak XX to the biexciton-to-exciton transition.\cite{Seguin2004, Rice2005} The large negative binding energies of XX ($> 10$~meV) are commonly observed in III-N QDs\cite{Schomig2004, Martin2005, Simeonov2008, Winkelnkemper2008, Bardoux2009, Sebald2011, Amloy2012, Chen2012}. It is due to the residual strain, even in dots with such small sizes, which enhances the repulsive exciton-exciton Coulomb interaction\cite{Simeonov2008}. Despite the residual strain, both the PL energy (Fig.~\ref{fig:pl_power_depend}(a)) and decay time (Fig.~\ref{fig:pl_power_depend}(d)) of the X-peak remain nearly constant as the excitation intensity increases significantly. These again supports that the X-peak comes from the discrete ground state in a zero-dimensional dot, and is thus un-affected by carrier screening. The carrier density only changes the relative occupation among different energy levels, leading to changes in the relative intensities of corresponding spectral lines.

The above study shows that the optical properties of the QW-ND, including PL energy, decay time and IQE, depend strongly on the excitation intensity as a result of strain and screening. On the other hand, in the QD regime, the effect of strain and screening on the exciton emission from the QD-ND becomes obscure due to the energy-level quantization.

\begin{figure*}
  \includegraphics{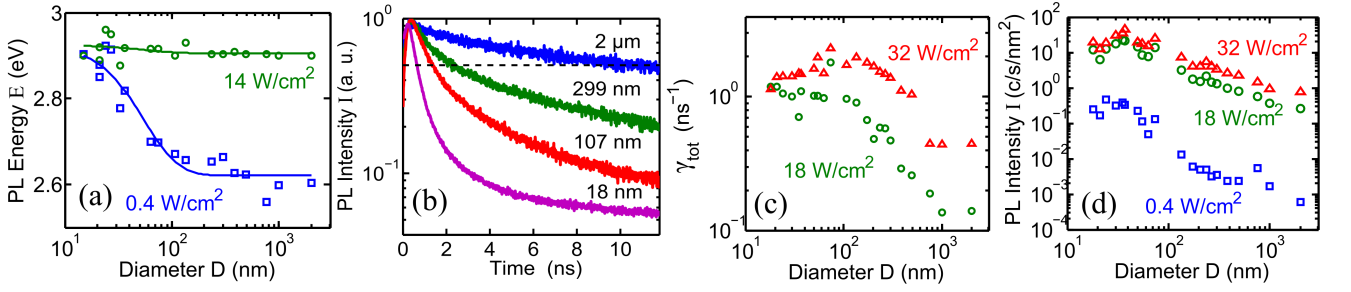}
	\caption{(a) The PL energy $E$ of single NDs vs. the diameter $D$ of the ND at excitation intensity $P=0.4$~W/cm$^2$ (blue square) and $14$~W/cm$^2$ (green circle). The single NDs are randomly chosen and isolated using a confocal microscope setup. The ND diameter is the average diameter of the ND's containing array which has a 2~nm standard deviation. The data are fitted using Equ.~\ref{equ:pl_energy} assuming the same $E_0$ and $\kappa$ for both low- and high-$P$ data, resulting in $E_0 = 2.92$~eV, $B_{m, low-P} = 303$~meV, $B_{m, high-P} = 19$~meV and $\kappa=0.055$~nm$^{-1}$.
(b) The TRPL decay traces of four ensembles of NDs with diameters $D=2~\mu$m (blue), 299 nm (green), 107 nm (red) and 18 nm (pink), respectively, at excitation intensity $P=18$~W/cm$^2$. Each ensemble contains a 6$\times$6 array of NDs.
(c) The total decay rate $\gamma_{tot} = 1/\tau_{tot}$ of ND ensemble vs. ND diameter $D$ at excitation intensities $P=18$~W/cm$^2$ (green circle) and $P=32$~W/cm$^2$ (red triangle). $\tau_{tot}$ is defined as the time taken for the PL intensity to decay from the maximum intensity to half maximum to avoid the complexity of stretched exponential decays.
(d) The PL intensity per unit InGaN area of ND ensemble vs. ND diameter $D$ at excitation intensities $P=0.4$~W/cm$^2$ (blue square), $P=18$~W/cm$^2$ (green circle) and $P=32$~W/cm$^2$ (red triangle).}
	\label{fig:efficiency_enhancement}
\end{figure*}

\label{QW to QD}
To obtain a quantitative comparison of the effect of strain relaxation on the optical properties of NDs, we performed a controlled systematic study of how PL energy, decay time, and intensity change with the ND diameter. We measured NDs of 21 different diameters, varying from $D=$15~nm -- 2~$\mu$m, fabricated on the same single-QW wafer of fixed thickness ($<2$ mono-layer fluctuation) and fixed indium fraction ($<2$\% fluctuation).
For each diameter, we measured one randomly chosen ND from a $6 \times 6$ array for PL energy study (Fig.~\ref{fig:efficiency_enhancement}(a)) or all 36 NDs for TRPL and PL intensity studies (Fig.~\ref{fig:efficiency_enhancement}(b-d)).

\label{- E vs. D}
To investigate strain relaxation from the QW to the QD regime, we examined the increase of PL energy $E$ with ND diameter $D$ reduction as a result of reduced built-in fields as shown in Fig.~\ref{fig:efficiency_enhancement}(a). At a high excitation intensity of $P = 14$~W/cm$^2$, $E$ increases by only $\sim30$~meV because of strong photo-carrier screening effects in large NDs. At a low excitation intensity of $P = 0.4$~W/cm$^2$ the photo-carrier screening effect is weak, so that the measured PL energy is closer to the intrinsic exciton transition energy without screening. As $D$ decreases from 2~$\mu$m to 100~nm, $E$ increases gradually by $\sim50$~meV; with further reduction of $D$ from $\sim100$~nm to 15~nm, $E$ increases rapidly by $\sim150$~meV. It shows that the effect of strain-relaxation is most significant for NDs of $D<100$~nm. 

The measured relation between PL energy $E$ and ND diameter $D$ can be described by a compact analytical equation\cite{Note1}:
\begin{equation}
E(D) = E_0 - B_m [1-\mbox{sech}(\kappa D/2)].
\label{equ:pl_energy}
\end{equation}
Here, $E_0$ corresponds to the exciton energy of a fully strain-relaxed InGaN ND, in the limit of $D=0$~nm; $B_m$ corresponds to the energy redshift from $E_0$ in fully strained large InGaN ND, in the limit of $D \rightarrow \infty$, and varies with the excitation intensity $P$; $1/\kappa$ corresponds to a strain-relaxation length with defines the region from the ND sidewall where strain relaxation is significant. Knowing $E_0$, $B_m$ and $\kappa$ we can also derive a phenomenological exciton potential profile along the ND radius\cite{Note1}. Fitting the data in Fig.~\ref{fig:efficiency_enhancement}(a), where a rapid shift of PL energy was observed for $D<100$~nm, we obtained $1/\kappa = 18$~nm, consistent with earlier numerical predictions\cite{Kawakami2010}. 

The reduction of strain in smaller NDs corresponds to an increase in the oscillator strength $f_{os}$, which should lead to an increase in the radiative decay rate $\gamma_r$ and IQE $\eta_{int}$. However, the excitation efficiency $\eta_{abs}$, LDPS $\rho$ and collection efficiency $\eta_{col}$ will also change with ND diameter. Taking into account these factors by numerical simulation, we can extract $f_{os}$, $\gamma_r$ and $\eta_{int}$ from the total decay time and PL intensity.

\label{- Tau and I vs. D}
We obtained the total decay time vs. diameter from the TRPL of ND ensembles. Fig.~\ref{fig:efficiency_enhancement}(b) shows TRPL traces of four ensembles. Clearly the decay accelerates as the diameter decreases. All TRPL curves are stretched exponential, as expected for QW-NDs as well as for QD-ND ensembles with inhomogeneous ND properties. We characterized their decay using the half-life-time $\tau_{tot}$, the time it takes to go to half of the initial PL amplitudes (dashed line in Fig.~\ref{fig:efficiency_enhancement}(b)). Fig.~\ref{fig:efficiency_enhancement}(c) shows the total decay rate $\gamma_{tot} = 1/\tau_{tot}$ vs. ND diameter $D$. At both excitation intensities of $P=18$ and $32$~W/cm$^2$, the total decay rate (including both radiative and nonradiative decays) increases drastically with the reduction of $D$ from $2~\mu$m to $100$~nm as expected, but plateaus or even slightly decreases with further reduction of $D$. 
Corresponding to the enhancement in the total decay rate, we measured a drastic increase in $I$, the integrated PL intensity per unit ND area, at a fixed excitation intensity (Fig.~\ref{fig:efficiency_enhancement}(d)). 

\label{- Abs and Col Efficiency}
The measured $I$ is determined by the excitation intensity $P$, absorption efficiency $\eta_{abs}$, IQE $\eta_{int}$ and collection efficiency $\eta_{col}$. Therefore, $\eta_{int}$ can be calcualted by:
\begin{equation}
\overline{\eta_{int}} \propto I/(P \overline{\eta_{abs}}\,\overline{\eta_{col}}),
\label{equ:eta_int_2}
\end{equation}
in which $\overline{\eta_{int}}$, $\overline{\eta_{abs}}$ and $\overline{\eta_{col}}$ are averaged $\eta_{int}$, $\eta_{abs}$ and $\eta_{col}$ values over the entire ND and all dipole polarizations\cite{Note1}. The same definition is used for $\overline{\gamma_r}$, $\overline{\gamma_{tot}}$, $\overline{F_p}$ and $\overline{f_{os}}$ in the rest of this work.
% For simplicity, all average-of-products are approximated using product-of-averages, which means that the amount of improvements in $\overline{\eta_{int}}$, $\overline{\gamma_r}$ and $\overline{f_{os}}$ we obtained in this work are order-of-magnitude estimations.

\begin{figure*}
  \includegraphics{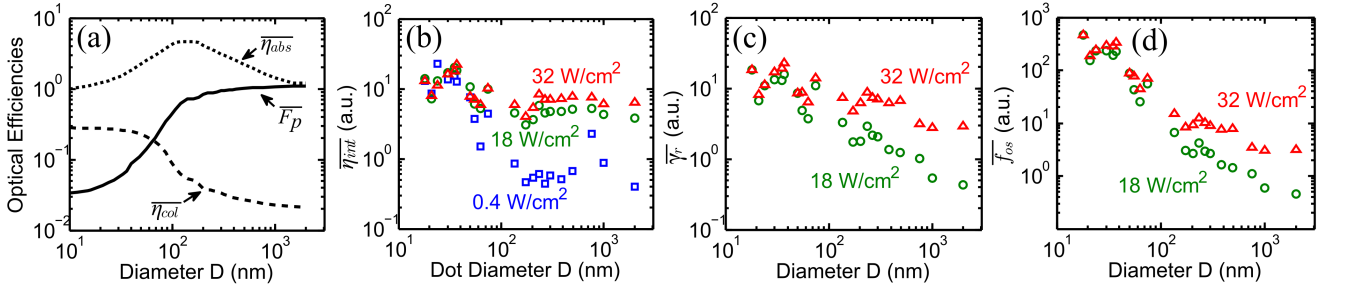}
	\caption{(a) FDTD simulation results\cite{Note1}: average absorption efficiency $\overline{\eta_{abs}}$ (dotted line), LDPS factor $\overline{F_p}$ (solid line) and collection efficiency $\overline{\eta_{col}}$ (dashed line) vs. ND diameter $D$. The $\overline{\eta_{abs}}$ curve is normalized to its value at $D=2$~$\mu$m and, hence, only shows the relative trend; while the $\overline{F_p}$ and $\overline{\eta_{col}}$ curves are absolute values. (b-d) Average IQE $\overline{\eta_{int}}$, radiative decay rate $\overline{\gamma_r}$ and oscillator strength $\overline{f_{os}}$ vs. $D$ extracted from Fig.~\ref{fig:efficiency_enhancement}(c) and (d) using Equ.~\ref{equ:eta_int_2}, \ref{equ:gamma_r_2} and \ref{equ:gamma_r}, respectively. The vertical axes of all three figures have arbitrary units (a.u.).}
	\label{fig:fdtd}
\end{figure*}

We evaluate $\eta_{abs}$ and $\eta_{col}$ using the finite-difference-time-domain (FDTD) method\cite{taflove2005computational}. The results are summarized in Fig.~\ref{fig:fdtd}(a).
The $\overline{\eta_{abs}}$ are nearly the same in the QW and QD limits.
The highest $\overline{\eta_{abs}}$ appears at $D\sim 150$~nm when the GaN nanopillar effectively forms a low-quality cavity in the lateral direction for light at 390~nm wavelength\cite{Note1}. The first-lens collection efficiency $\overline{\eta_{col}}$ increases from merely 2\% in the QW limit to about 30\% in the QD limit; and the most drastic increase occurs in the region 40~nm $<D<100$~nm. A similar trend in $\eta_{col}$ has been predicted for a dipole emitter embedded near the end of a semi-infinite dielectric nanowire\cite{Friedler2009, Bleuse2011}. It was shown that\cite{Friedler2009}, for a large nanopillar, most light is coupled into two guided HE$_{11}$ modes, one propagating upwards and the other downwards into the substrate. However, the upwards-propagating guided mode is mostly reflected back into the substrate by the top facet of the nanopillar. For a small nanopillar\cite{Bleuse2011}, the coupling to the guided mode is strongly suppressed while the dissipation into free-space modes is relatively enhanced, leading to an increased collection efficiency. 

\label{-IQE}
Using Equ.~\ref{equ:eta_int_2} and the calculated $\overline{\eta_{abs}}$ and $\overline{\eta_{col}}$, we extracted $\overline{\eta_{int}}$ as shown in Fig.~\ref{fig:fdtd}(b). Note that we cannot obtain the absolute value of $\overline{\eta_{int}}$ from Equ.~\ref{equ:eta_int_2}, so $\overline{\eta_{int}}$ is a relative value with an arbitrary unit (a.u.). Nonetheless, Fig.~\ref{fig:fdtd}(b) suggests that the $\overline{\eta_{int}}$ in the QD limit is enhanced compared to that in the QW limit; and the enhancement saturates or even slightly decreases when $D$ reduces further. The enhancement is $\sim 10$ folds at $P = 0.4$~W/cm$^2$, but only $\sim50$\% at $P=18$ and 32~W/cm$^2$. This is because at high $P$ the oscillator strength $f_{os}$ in large NDs is already enhanced due to screening. 
The saturation of $\overline{\eta_{int}}$ at $D<40$~nm in Fig.~\ref{fig:fdtd}(b) could be due to two factors: the radiative decay rate $\overline{\gamma_r}$ enhancement is saturated, and at $D = 40$~nm the $\overline{\eta_{int}}$ is already close to 100\%.

\label{- Radiative Rate}
The variation of $\overline{\gamma_r}$ with $D$ can be obtained from $\overline{\eta_{int}} = \overline{\gamma_r}/\overline{\gamma_{tot}}$ as follows:
\begin{equation}
\overline{\gamma_r} \propto I \, \overline{\gamma_{tot}}/(\overline{\eta_{abs}}\,\overline{\eta_{col}}P),
\label{equ:gamma_r_2}
\end{equation}
in which $\overline{\gamma_{tot}}$ is measured in Fig.~\ref{fig:efficiency_enhancement}(c).
Fig.~\ref{fig:fdtd}(c)shows that the $\overline{\gamma_r}$ improves by only $\sim10$ folds as $D$ decreases from $2~\mu$m to 40~nm and saturates at $D<40$~nm. 
However, the 10-fold improvement is not the true potential of $\overline{\gamma_r}$ improvement in our NDs. According to Fermi's Golden Rule\cite{Dirac1927, novotny2006principles}, $\gamma_r$ is determined by both the oscillator strength $f_{os}$ and LDPS $\rho$:
\begin{equation}
\overline{\gamma_r} \propto \overline{f_{os}} \, \overline{F_p},
\label{equ:gamma_r}
\end{equation}
in which, $F_p = \rho/\rho_0$ is the ratio of the LDPS $\rho$ of a dipole in a nanopillar to the LDPS $\rho_0$ of the same dipole in bulk GaN. Note that the oscillator strength $f_{os}$, defined as the ratio between the radiative decay rate of the ND in bulk GaN and that of a classical electron oscillator in bulk GaN\cite{siegman1986lasers}, is proportional to the square of the radiative transition matrix element\cite{yu2010fundamentals}. As we shall show below, the $\overline{\gamma_r}$ in the QD limit is mainly limited by the LDPS factor $F_p$.

\label{- Purcell Factor}
The $F_p$ changes with the change of local dielectric environment. As shown in Fig.~\ref{fig:fdtd}(a), in the QD limit, the average LDPS factor $\overline{F_p}$ is strongly suppressed, with $\overline{F_p}< 1/10$. Such a strong suppression in the LDPS should be universal for all nano-emitters closely surrounded by air-dielectric interfaces, such as colloidal QDs\cite{Murray2000}, QDs in nanowires\cite{Deshpande2013, Bounouar2012, Tatebayashi2012, Holmes2013} and QDs at the apices of micro-pyramids\cite{Baier2004}. Experimentally, it has been demonstrated for nano-emitters in nano-spheres\cite{Schniepp2002} and nanowires\cite{Bleuse2011}. The radiative decay rate $\gamma_r$ in our NDs in the QD limit can be further enhanced by increasing the LDPS. For example, simply by conformal-coating the sample with 150~nm GaN, we can recover the $\overline{F_p}$ to unity while maintaining the relatively high $\overline{\eta_{col}} = 15$\% in QD-NDs\cite{Note1}. Larger $\overline{F_p}$ and $\overline{\eta_{col}}$ enhancements may be achieved with more sophisticated structures, such as by enclosing a ND in a micro-cavity\cite{Gazzano2013} and in a tapered nanowire\cite{Claudon2010}.

\label{- Dipole Moment}
With the results of $\overline{\gamma_r}$ and $\overline{F_p}$ we can calculate the oscillator strength $\overline{f_{os}}$ using Equ.~\ref{equ:gamma_r}. As shown in Fig.~\ref{fig:fdtd}(d), $\overline{f_{os}}$ is enhanced by over 100 folds in the QD limit compared to the QW limit at $P = 18$ and 32~W/cm$^2$. The $\overline{f_{os}}$ enhancement is a direct result of the strain-relaxation-induced reduction in the piezo-electric polarization fields, which leads to a better overlap between the electron and hole wavefunctions.

\label{Conclusion}
In conclusion, we have systematically investigated the optical properties of individual and ensembles of InGaN/GaN nanodisks with precisely controlled diameters varying from the QD limit of $D$ less than 40~nm, to the QW limit of $D$ up to 2~$\mu$m. We found significant strain relaxation in nanodisks with diameters less than 100~nm, leading to a 100-fold enhancement in the oscillator strength in the QD limit compared to in the QW limit. Together with the 10-fold suppression in the local density of photon states, this leads to a 10-fold enhancement in the radiative decay rate, which can be further enhanced by increasing the local density of photon states. 

We acknowledge financial supports from the National Science Foundation (NSF) under Awards ECCS 0901477 for the work related to materials properties and device design, ECCS 1102127 for carrier dynamics and related time-resolved measurements, and DMR 1120923 (MRSEC) for work related to light-matter interactions. The work related to epitaxial growth, fabrication, and photon antibunching properties were also partially supported by the Defense Advanced Research Project Agency (DARPA) under grant N66001-10-1-4042. Part of the fabrication work was performed in the Lurie Nanofabrication Facility (LNF), which is part of the NSF NNIN network.

%\bibliography{References}
%\bibliographystyle{apsrev4-1}


\end{document}